# A 6bit, 1.2GSps Low-Power Flash-ADC in 0.13μm Digital CMOS

Christoph Sandner, Martin Clara, Andreas Santner, Thomas Hartig, Franz Kuttner

Infineon Technologies Austria, Development Center Villach
A-9500 Villach, Siemensstr. 2
mailto:christoph.sandner@infineon.com

**Abstract**

*A 6bit flash-ADC with 1.2GSps, wide analog bandwidth and low power, realized in a standard digital 0.13μm CMOS copper technology is presented. Employing capacitive interpolation gives various advantages when designing for low power: no need for a reference resistor ladder, implicit sample-and-hold operation, no edge effects in the interpolation network (as compared to resistive interpolation), and a very low input capacitance of only 400fF, which leads to an easily drivable analog converter interface.*

*Operating at 1.2GSps the ADC achieves an effective resolution bandwidth (ERBW) of 700MHz, while consuming 160mW of power. At 600MSps we achieve an ERBW of 600MHz with only 90mW power consumption, both from a 1.5V supply. This corresponds to outstanding Figure-of-Merit numbers (FoM) of 2.2 and 1.5pJ/convstep, respectively. The module area is 0.12mm².*

## 1. Introduction

Flash Analog-to-Digital Converters (ADCs) are still the architecture of choice, where maximum sample rate and low to moderate resolution is required. A typical example is e.g. the read-write channel of a disk drive system, where customers often ask for the maximum sample rate that is offered by the currently available technology generation.

However, there are additional applications in the wireless area coming up, where a flash-ADC gives sufficient accuracy at the required large analog bandwidth, for example in ultra-wideband (UWB) systems. Since a lot of wireless applications are hand-held as well, this poses an important constraint to the specification of the ADC, which is power consumption. The ADC can be the dominant block in terms of power consumption for the whole analog frontend of such a system.

In this paper we describe the concept, design and measurement results of a 6bit flash-ADC, optimized for low power at sufficiently high data rates and analog bandwidth. With 1.2GSps and ERBW of 700MHz the ADC fulfils the requirements currently under discussion for multi-band OFDM UWB systems [1]. Due to the large ERBW different receiver topologies, like low-IF or zero-IF can be supported, both at Nyquist rate, and under sub-sampling operation. However, the presented ADC architecture is not limited to these applications.

Although almost all recently published high-speed full-flash-ADCs employ resistive interpolation and averaging, the capacitive interpolation structure with distributed front-end sample-and-hold seems to have advantages in terms of power and area, even at GSps-speed. The outstanding FoM numbers prove the efficiency of the implemented architecture, when compared to state-of-the-art 6bit flash ADCs.

## 2. Converter Architecture

Averaging is a well known technique to improve the linearity of a flash-type ADC beyond the matching limit of the single comparator [2]. Multiple gain-stage architectures allow the use of interpolation to reduce the number of front-end amplifiers and thus the input capacitance of the converter. By scaling the amplifiers in the analog preprocessing chain from front to back also the overall power consumption can be optimized under the given gain/bandwidth constraints. There exist basically two different strategies for implementing a flash

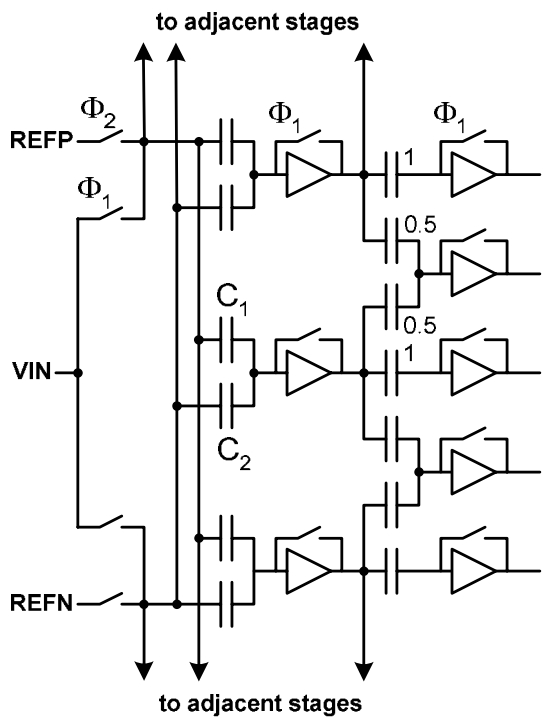

**Fig. 1: ADC frontend portion using capacitive interpolation**

ADC employing interpolation and averaging.

Resistive interpolation employs averaging resistors between the outputs of adjacent amplifiers. [3]. A common problem of such architectures is the need for over-range comparators to maintain linearity at the edges of the conversion range. Special circuit techniques [4] allow reducing the number of overrange comparators, but they rely on matching the termination resistor with the output resistance of the overrange blocks. Furthermore, an external sample-and-hold circuit is almost always required, which consumes a significant fraction of the total power budget in wide-band applications.

Capacitive interpolation, on the other hand, uses a purely reactive averaging network between the outputs of adjacent amplifiers [5]. A big advantage of capacitive interpolation is that it requires neither power consuming overrange comparators nor any static averaging termination. Also, no external sample-and-hold is required, because the interpolation capacitors at each stage are readily used as sampling capacitors implementing a multi-stage input-offset-sampling (IOS) architecture with distributed front-end sample-and-hold.

The principle of capacitive interpolation combined with distributed front-end sample-and-hold is illustrated in Fig. 1, showing a portion of the converter input. The inverting first stage amplifiers drive the interpolating capacitive network of the second stage, here implementing an interpolation factor 2. The interpolation factor at the converter input is given by the total number of front-end amplifiers minus 1. The total input capacitance of the converter in the sampling phase is given by the sum of the front-end capacitors plus wiring parasitics.

One drawback of the capacitive interpolation structure is the capacitive divider formed by the sampling capacitor and the input capacitance of the amplifier during the amplification phase. The overall gain of each stage is given by the intrinsic gain of the amplifier and the capacitive divider ratio. In order to minimize the total input capacitance of the converter, the sampling capacitors will be chosen as small as possible, for a resolution of 6 bits ultimately limited by capacitor mismatch. For minimum gain loss the amplifiers' input devices should therefore be as small as possible. Here, IOS helps to further reduce the input devices' effective mismatch, thus allowing additional downscaling to save power and area.

## 3. Implemented Design

The block diagram of the implemented A/D-converter is shown in Fig. 2. The circuitry is fully differential, although drawn single ended for simplicity. The input signal is sampled in the first clock phase $\phi_1$ by the front-end amplifiers forming a distributed sample-and-hold, while the subsequent stages sample the offset voltage. During the second clock phase $\phi_2$ the reference voltage is applied to the bottom plates of the front-end sampling capacitors and the difference between input voltage and reference voltage is amplified before it is latched by the comparators at the end of this clock phase.

The capacitive load at each amplifier output and thus the bandwidth of the amplifier is linearly related to the interpolation factor. To optimize the power-bandwidth product of the amplifier, while still profiting from the averaging property, the minimum possible interpolation factor of 2 is chosen at the output of each gain stage. To reach 64 decision levels after three interpolating stages, an interpolation factor of 8 is implemented at the converter input, thus formed by 9 parallel input amplifiers. A fourth gain stage directly drives the latching comparator, leading to an 8-2-2-2-1 interpolation topology. Since the overall interpolation factor is 64, only two reference voltages are needed at the input of the 6 bit ADC, thus avoiding the silicon area and power penalty of a low-resistance reference ladder. Instead, the reference voltages are generated by capacitive voltage division during the amplification phase $\phi_2$ (referring to single-ended representation in Fig. 1):

$$V_{REFi} = V_{REFN} + \frac{C_1}{C_1 + C_2}(V_{REFP} - V_{REFN}) \quad (1)$$

Through optimized sizing of the converter front-end, it was possible to reach a very low input capacitance of only 400fF. The subsequent gain stages are progressively scaled down by a factor of 2, thus achieving a very compact layout, since all 4 gain stages end up to have the same height (Fig. 4).

Regarding DC performance, the input referred offset is the most important specification. Since, due to minimum sizing for maximum speed, the offset of the final comparator stage is dominant, this offset must be reduced by achieving sufficient gain in the preamplifiers, thus reducing the input referred offset. The amplifier block (Fig. 3) consists of a differential pair with resistive load. PMOS switches connected between inputs and outputs provide the offset sampling. The gain of each stage is chosen to be 2.5 for achieving maximum bandwidth.

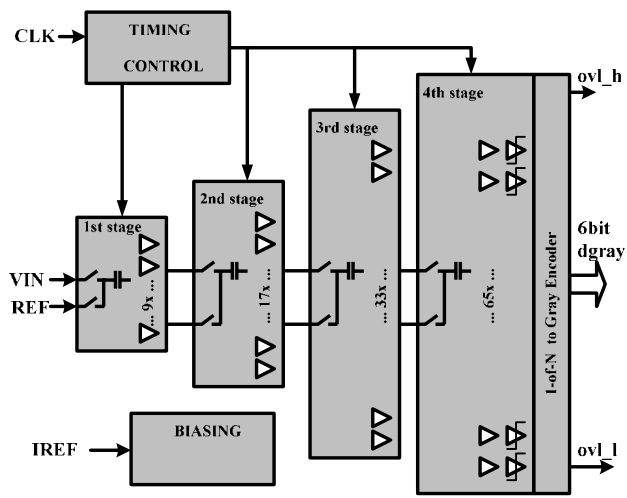

Fig. 2: Implemented ADC Block Diagram

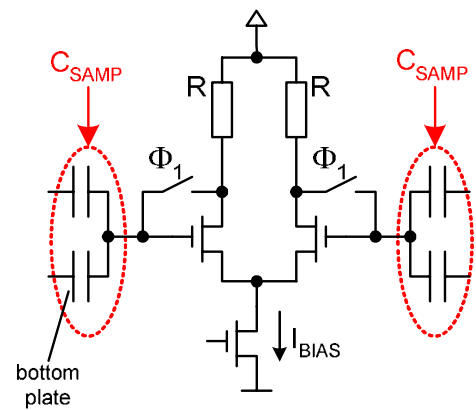

Fig. 3: Amplifier Block Diagram

The final comparator latch is shown in Fig. 5. The input voltage difference is first converted into a current difference and then fed to the cross-coupled latch, formed by transistors N2. During the sampling phase, when the amplifiers sample the analog input signal, the CLK signal is high, thus keeping the comparator outputs OUTP and OUTN at the same level. During amplification phase CLK goes low, thus releasing the outputs, and the latch can decide whether the input voltage difference was positive or negative. The use of a single NMOS clock switch between both outputs is not possible, due to the low supply voltage of 1.5V. The *gm* of transistors N2 must be chosen such that the desired Bit Error Rate (BER) specifications are met. Main advantages of this topology are the almost rail-to-rail output swing, and the reduced kick-back to the pre-amplifiers due to the current mirrors.

For increasing the BER the 64 outputs of the comparator row are again latched twice by differential latches, before entering a first order bubble correction stage, which also converts the thermometer code to a 1-out-of-N code. Finally, this code is converted into a 6 bit Gray-code using a ROM-table based on a current steering topology optimized for high frequency operation. For test purposes the digital data is downsampled by a factor of 64, thus standard CMOS pads can be used for the digital outputs.

Fig. 4 shows the layout plot of the implemented test-chip. It was fabricated in a 0.13μm digital CMOS technology with standard-threshold MOS devices (no low-VT option), single poly, 4 thin copper metals, and 2 thick copper metals. The capacitors are of metal-metal sandwich type, taking advantage of both vertical and lateral capacitances.

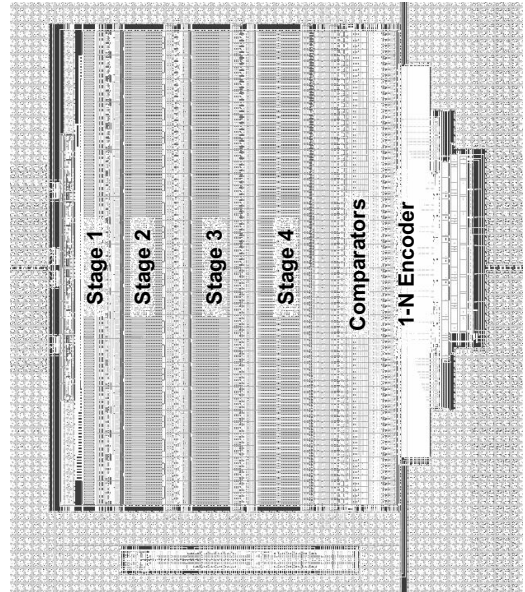

**Fig. 4: ADC Layout**

## 4. Measurements

All measurements are done on a standard PCB with ceramic TQFP-44 package and no socket, at room temperature and nominal supply of 1.5V. Both, analog input and sampling clock are applied differentially to the chip by using baluns.

Fig. 6 shows the DNL of the converter, measured at 600MSps with a full-scale 50MHz sine-wave input employing a histogram method. The measured peak DNL of 0.4LSB thus already includes dynamic effects at that sampling rate. Peak INL is measured <0.6LSB.

In Fig. 7 the maximum sampling rate of the ADC is explored for nominal bias current. The analog signal frequency is kept constant at 121MHz, while the sampling clock frequency is varied on the x-axis. The ADC shows 5bit ENOB up to a sample rate of 1.4GSps. The clock frequency limit for the digital circuitry is around 1.6GSps.

Fig. 8 shows the SNR and SNDR for a sample rate of 600MSps, at reduced bias current to save power. The SNDR at 51MHz is 35.5dB, with an SFDR of 52dB and a THD of 49dB, thus proving the excellent linearity of the capacitive interpolation topology. The SNDR drops by 3dB at an ERBW of 600MHz.

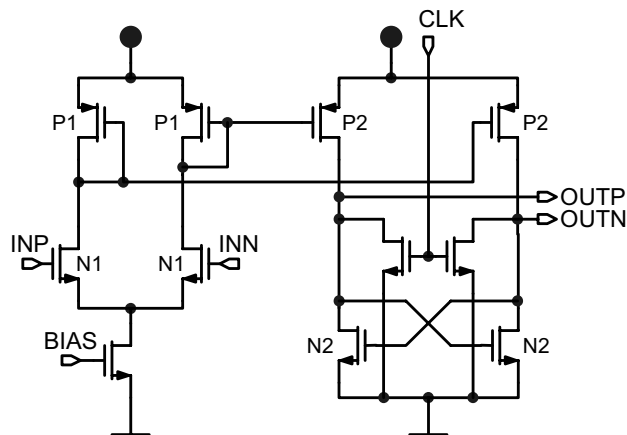

**Fig. 5: Final Comparator Latch**

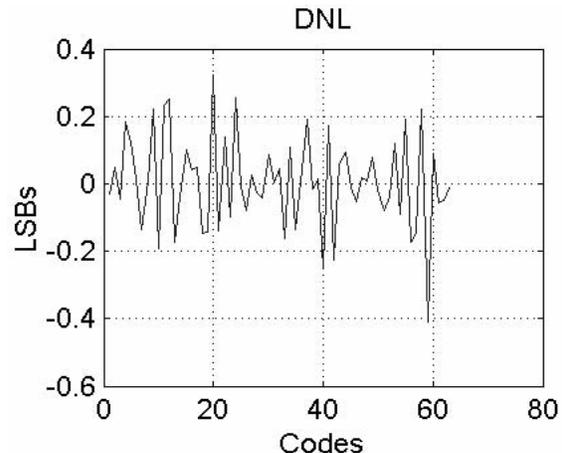

**Fig. 6: Measured DNL typ. <0.4LSB   (at 600MSps)**

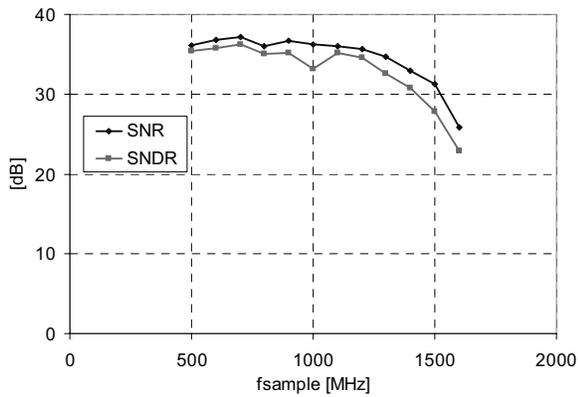

**Fig. 7: Measured Dynamic Performance vs. fsample (fsig = 121MHz)**

In Fig. 9 the same measurement is done for 1.2GSps at nominal bias current. SNDR starts at 35.8dB for low frequencies (THD is 46dB), dropping by 3dB at an ERBW of 700MHz. These measurements show that this ADC topology is very well suited for applications with wide analog bandwidth.

To do a comparison with state-of-the-art 6bit flash-ADCs, a Figure-of-Merit (FoM) is calculated [8]:

$$FoM = \frac{Power}{2^{ENOB,DC} \cdot 2 \cdot ERBW} \quad [pJ/convstep] \quad (2)$$

For our ADC we achieve a FoM of 2.2pJ/conv at 1.2GSps, and 1.5pJ/conv at 600MSps. As can be seen in Fig. 10 these are the best FoM numbers for flash ADCs ever published [4,6-9]. Although a smaller feature size technology is used in this work, the achieved performance nevertheless proves the efficiency of the capacitive interpolation architecture with distributed sample-and-hold for flash ADCs in the GHz range.

## 5. Acknowledgements

We thank C. Kropf for layout work, P. Schreilechner for board design, P. Bogner, D. Draxelmayr and G. Knoblinger for fruitful discussions.

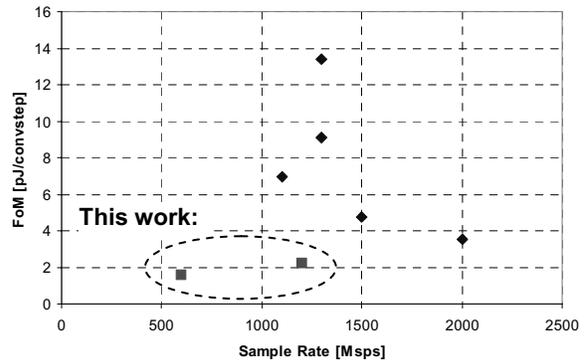

**Fig. 10: Comparison to state-of-the-art 6b Flash ADCs**

## 6. References

[1] A. Batra et al, "Multi-band OFDM Physical Layer Proposal", IEEE 802.15 WPAN High Rate Alternative PHY Task Group 3a (TG3a), Sep. 2003, http://www.ieee802.org/15

[2] Klaas Bult, Aaron Buchwald, "An Embedded 240-mW 10-b 50-MS/s CMOS ADC in 1-mm²", *IEEE Journal of Solid-State Circuits, Vol. 32, No. 12*, Dec. 1997, pp. 1887-1895.

[3] Kevin Kattmann, Jeff Barrow; "A technique for reducing differential non-linearity errors in flash A/D converters", *IEEE International Solid-State Circuits Conference, Vol. XXXIV*, Feb. 1991, pp. 170 – 171.

[4] Peter Scholtens, Maarten Vertregt, "A 6-b 1.6-Gsample/s Flash ADC in 0.18-um CMOS Using Averaging Termination", *IEEE Journal of Solid-State Circuits, Vol. 37, No. 12,* Dec. 2002, pp. 1599-1609.

[5] Keiichi Kusumoto, Akira Matsuzawa and Kenji Murata, "A 10b 20MHz 30mW Pipelined Interpolating CMOS ADC", *IEEE Journal of Solid-State Circuits, Vol. 28, No. 12*, December 1993, pp. 1200-1206

[6] Michael Choi, Asad A. Abidi, "A 6-b 1.3-Gsample/s A/D Converter in 0.35μm CMOS", *IEEE Journal of Solid-State Circuits, Vol. 36, No. 12*, Dec. 2001, pp.1847-1858.

[7] Govert Geelen, "A 6b 1.1GSample/s CMOS A/D Converter", *IEEE International Solid-State Circuits Conference, Vol. 44*, Feb. 2001, pp. 128-129.

[8] Xicheng Jiang, Zhengyu Wang, M. Frank Chang, "A 2GS/s 6b ADC in 0.18μm CMOS", *IEEE International Solid-State Circuits Conference*, Vol. 46, Feb. 2003, pp. 322-323.

[9] Koen Uyttenhove, Michiel Steyaert, "A 1.8-V 6-bit 1.3-GHz flash ADC in 0.25μm CMOS", *IEEE Journal of Solid-State Circuits, Vol. 38, No. 12,* July 2003, pp. 1115-1122.

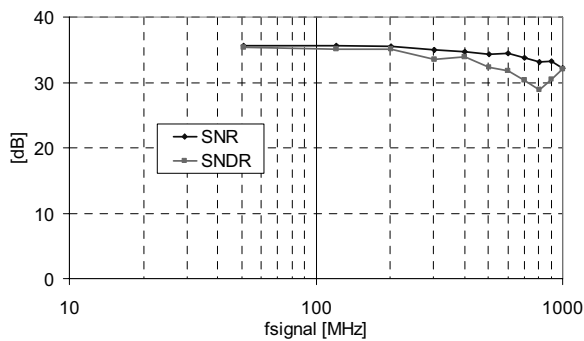

**Fig. 8: Measured Dynamic Performance vs. fsignal at 600MSps**

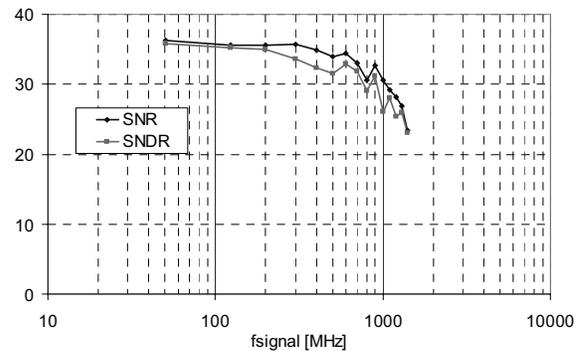

**Fig. 9: Measured Dynamic Performance vs. fsignal at 1.2Gsps**